# Magnetite: Raman study of the high-pressure and low-temperature effects.

L.V. Gasparov, D. Arenas

Department of Chemistry and Physics, University of North Florida,

St. John's Bluff Rd. South, Jacksonville, Fl 32224, USA

K. -Y. Choi, G. Guentherodt

2. Physikalisches Instituit, RWTH-Aachen, 52056 Aachen, Germany

H. Berger, L. Forro, G. Margaritondo

*EPFL, Lausanne, Switzerland*

V. V. Struzhkin, R. Hemley

Geophysical Laboratory, Carnegie Institution of Washington, Washington DC, USA

## Abstract

We report the results of a low-temperature (300K-15K) high-pressure (up to 22GPa) Raman study of the Verwey transition in magnetite ($Fe_3O_4$). We use additional Raman modes observed below the Verwey transition to determine how the transition temperature changes with the quasihydrostatic pressure. Increase of the pressure results in the linear decrease of the Verwey transition temperature, with no discontinuity. The corresponding pressure coefficient $dT_V/dP$ is found to be -5.16±1.19 K/GPa. Such a decrease is substantially larger than the one predicted by the mean-field Coulomb interaction model of the transition.

Fe$_3$O$_4$ (magnetite) was the first magnetic material known to mankind and the earliest compound known [1] to manifest the charge-ordering transition discovered by Verwey in 1939. Magnetite has recently attracted much of attention [2, 3] on the account of its charge carriers strong spin polarization at the Fermi level. This compound has a potential to become one of the leading materials for spintronics [3].

Fe$_3$O$_4$ was extensively studied for more than sixty years. Yet the nature of its Verwey transition is still a puzzle. At ambient pressure this is a first-order transition at T$_V$ = 120 K, with changes of the crystal structure, latent heat, and a decrease of the dc-conductivity by two order of magnitude. There are several competing theories of the transition [4]. However, none of them is capable to describe the entire body of experimental data. Recently, Brabers et al. [5-7] suggested a mean filed description of the transition based on an effective interionic Coulomb potential. This model yields dT$_v$/dp = -2.76 K/GPa, a value that can be verified by a variety of experimental techniques. Indeed several groups [8-13] reported high-pressure transport (except Ref. 11) measurements of T$_V$. The majority of reports [8-11] indicate that T$_V$ follows a linear pressure dependence though, with different slopes. The transport measurements of Refs. 9, 10, and 12 yielded values of dT$_v$/dp close to that predicted by Brabers et al. However, Ref. 8 and 11 report values closer to –5K/GPa. It is important to note a basic disadvantage of transport measurements related to the fact that transport properties are always governed by the interplay of carrier concentration, which is a function of the density of states, and mobility, which is a function of quite a few parameters including defect concentration. In order to overcome this problem, one could use a measurement technique capable of detecting structural changes associated with the transition such as optical spectroscopy: this has been our motivation to employ Raman spectroscopy for the study of magnetite.

The magnetite single crystals were grown by a chemical vapor transport technique using stoichiometric $Fe_3O_4$ microcrystalline powder obtained by reduction reaction of ferric oxide ($Fe_2O_3$). This procedure yielded near-stoichiometric single crystals with typical size of 4 x 4 x 1 mm. X-ray diffraction confirmed the spinel-type structure of the crystals. Transport measurements found the abrupt increase in resistance below T=120K, characteristic of the Verwey transition.

Raman measurements at atmospheric pressure were carried out on the freshly cleaved surfaces of the as-grown single crystals using Dilor XY-modular triple spectrometer equipped with a liquid nitrogen cooled CCD detector. The incident laser power on the sample did not exceed 15 mW. The sample temperature was maintained in He-bath cryostat at controlled temperatures from 5 to 300K.

High-pressure unpolarized Raman spectra were measured with a Jobin Ivon HR 460 single stage monochromator equipped with CCD. Low temperature was assured by the He-flow cryostat. The magnetite sample of about 15μm was placed in the center of a gasket together with the small ruby crystals necessary to measure the applied pressure. The gasket with the sample and ruby crystal was positioned between two synthetic diamond anvils. The opening was filled with neon gas serving as a pressure transfer medium and locked with the help of specially constructed system of nuts and sealing. Pressure was applied to the anvils and transferred nearly hydrostatically to the sample. The laser light was shone through the diamonds on the sample and Raman scattering was collected. The pressure was measured by the shift in frequency of the ruby luminescence line. While using neon as a pressure transfer medium, we discovered a substantial overheating of the sample. To reduce this problem, we had to defocus the laser. Furthermore, we put the sample in direct contact with the diamond anvil and used NaCl as the pressure transfer medium. Such an approach resulted in about 2 GPa pressure gradient over the different areas of the sample, corresponding to an experimental uncertainty of ±2GPa.

The effects of the Verwey transition on the Raman spectra of magnetite at ambient pressure have been addressed in a number of publications [14-17]. Above the Verwey transition (T>120 K), magnetite has a cubic inverse-spinel structure [18] belonging to the space group $O^7_h$ (Fd3m). Group theory predicts five Raman-active modes. In spite of years of x-ray and neutron studies of magnetite, there is no consensus on its exact low temperature structure. However, there is no doubt that below the Verwey transition magnetite has much bigger unit cell [19, 20] leading to a dramatic increase in the number of phonon modes.

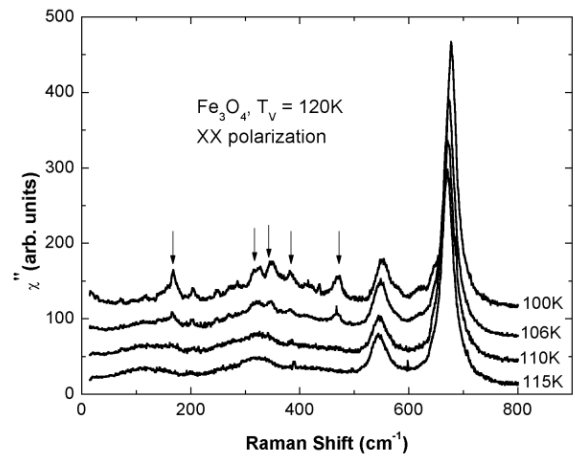

Fig.1. Ambient pressure polarized Raman spectra of the magnetite sample. The spectra are shifted vertically in order to more easily see the effect of the transition. Arrows indicate phonon modes that can serve as the markers of the Verwey transition. The marked temperature is the value read by the sensor. We estimate the actual temperature to be about 10K higher.

In Fig.1, we plot imaginary part of the Raman response function $\chi''$, which is obtained by dividing the Raman intensity by the corresponding Bose-factor. Below the transition we observe 13 Raman modes. From these modes four are present above the Verwey transition (metallic phase) whereas the modes at 165, 206, 290, 320, 350, 390, 470, 630 cm$^{-1}$ are present only in the semiconducting phase. The Raman modes at 165, 290, 320, 350 and 470 cm$^{-1}$ can serve as convenient markers for the transition, Fig.1. From the appearance of these modes one can determine $T_V$ and study how different parameters affect it. The estimated experimental error for this procedure is ±10K.

Figures 2 and 3 illustrate our high-pressure Raman experiment. Fig.2a shows Raman spectra taken at ~20 GPa. As the temperature decreases, marker- modes appear in the spectrum revealing the transition to the metallic phase. Fig. 3(b) shows spectra taken at 60K but at different pressures. Similarly to Fig. 2(a), marker-modes appear in the spectrum taken at lower pressure indicating that the Verwey transition took place.

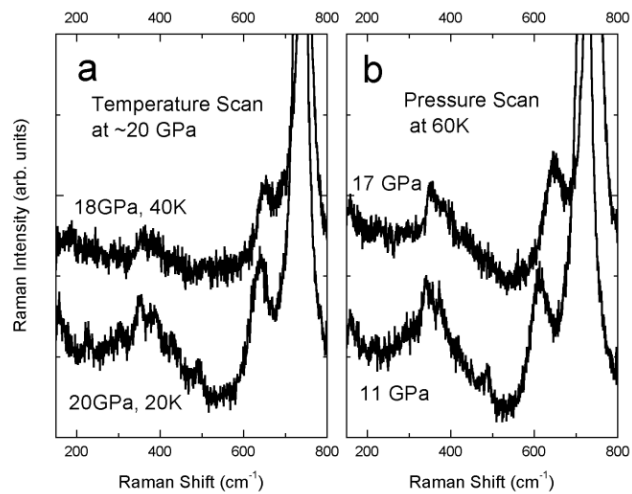

Fig.2 Unpolarized Raman spectra of magnetite measured at different pressures and temperatures. The spectra are again vertically shifted. Arrows indicate the modes characteristic of the semiconducting phase.

Fig. 3 shows the Verwey transition temperature as a function of the applied pressure. Within our experimental uncertainly, we do not observe any discontinuity in the $T_V$ dependence as reported in Ref. 12. However, we do observe the Verwey transition at pressures higher than 8GPa in contrast to Ref. 13. The critical temperature $T_V$ decreases linearly with pressure with a rate of −5.16±1.19 K/GPa, close to the slope reported in Refs. 8 and 11 and about twice the slope of Ref. 9,10 and 12. This dependence, if extrapolated to zero temperature, would result in the metal-semiconductor transition at 0K and ~25GPa, consistent [4] with a polaron-based mechanism of the transition.

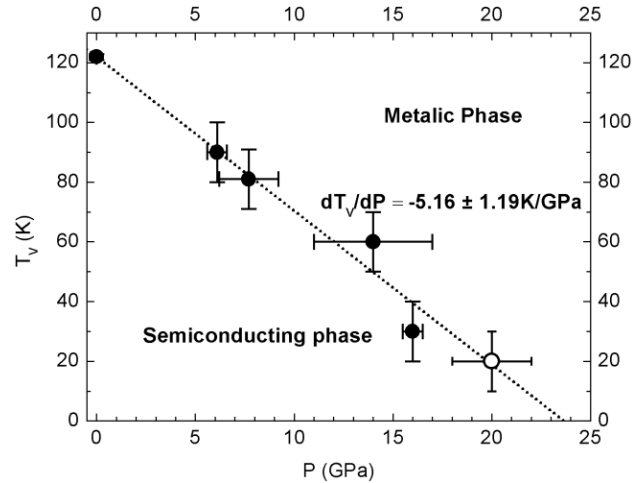

Fig.3. Raman data on the pressure effect on the Verwey transition. The dashed line is a linear fit to the data. The rather large error bars for the pressure are due to the pressure gradient since the transition occurs in only part of the sample. In particular, the point at 60K was obtained from three spectra during decompression. The point at 20 GPa (open circle) corresponds to the neon pressure medium. The other points were obtained with NaCl as pressure transfer medium.

In conclusion, we successfully used Raman spectroscopy to study how the temperature of the Verwey transition changes with the hydrostatic pressure. We observed a linear decrease of the transition temperature as the pressure increases. The corresponding pressure coefficient of $T_V$ was found to be $-5.16\pm1.19$ K/GPa, which is in contradiction with that predicted by Brabers et al. Our data extrapolated to higher pressure correspond to a metal semiconductor transition at 0K and 25GPa, consistent with a polaron-based scenario. We suggest that the discrepancy between pressure coefficients obtained from spectroscopic and transport measurements could be due to different temperature behavior of the mobility and the carrier concentration at the Verwey transition. These effects do not affect Raman data.

We thank D. Tanner and E. Ya. Sherman for valuable discussions. This work was supported by the Alexander von Humboldt Foundation, Research Corporation Cottrell College Science award #CC5290 and CC6130, Petroleum Research Fund award #40926-GB10, the Swiss Fonds National de la Recherche Scientifique and the EPFL. Daniel Arenas acknowledges the support of the UNF Undergraduate research award